\documentclass[12 pt, onecolumn,showpacs,preprintnumbers,amsmath,amssymb]{revtex4}
\usepackage{graphicx}

\usepackage{dcolumn}
\begin{document}

\title{Effects of three-body scattering processes on BCS-BEC Crossover }
\author{Raka Dasgupta}\email{rakadasg@bose.res.in}
% \author{J.K. Bhattacharjee}\email{jkb@bose.res.in}
\affiliation{S.N.Bose National Centre For Basic Sciences,Block-JD,
Sector-III, Salt Lake, Kolkata-700098, India}

\date{August 4, 2010}
\begin{abstract}
We investigate the BCS-BEC crossover taking into account an additional three-body interaction, which is essentially the scattering between the Cooper pairs and the newly formed bosons. We show that if the two-body interaction is attractive, the presence of this additional three-body term makes the crossover process a non-reversible one. Starting from a stable BEC state, crossover to BCS can be achieved; but if BCS state be the starting point, instead of a stable BEC region, what the system goes over to is a metastable condensed state.

\end{abstract}


\maketitle
\section*{I. INTRODUCTION}
The realization of tunable inter-atomic interactions via Feshbach resonances has made it possible to achieve the crossover from weak coupling BCS superfluidity to Bose-Einstein condensation
BEC of bound diatomic molecules. Recent experiments have successfully explored the crossover regime by means of studying the cloud size \cite{Bart}, expansion energy \cite{bourd}, resonance condensation \cite{grein,reg} and condensed nature of the fermionic atom pairs \cite{zw1, zw2}. As for the theoretical predictions,
BCS-BEC crossover was first addressed way back in 1969, in the seminal work of
Eagles \cite{eagl}. Later, using a variational prescription, Leggett \cite{leggett} showed 
that as the coupling strength is increased, the superconducting BCS ground state at zero temperature smoothly evolves into a BEC state of tightly bound molecules. 
Nozieres and Schmitt-Rink \cite{noz} and M. Randeria \cite{rand1} extended the analysis to a finite temperature. Since then, various aspects of the crossover problem has been widely investigated over the years \cite{ rand2,engel,melo,rann,levin1,levin2,ohashi,holland, fuchs, manini,strinati1, sheehy}.

Almost all the works have addressed the crossover phenomenon as a two-body scattering problem, where the interaction between two fermions has played the pivotal role. However, as Milstein et al. pointed out \cite{milst} it would be interesting to extend the approach to incorporate the effect of
higher order interactions in the crossover region. In fact, higher order scatterings and non linear interactions are being investigated in other domains of ultracold atom physics as well, starting from cubic interactions in BEC \cite{carr,gammal, mei}, to atom-dimer scattering in fermi systems \cite{iskin, petrov, mora} and also induced interactions in three-component Fermi gases \cite{torma}, and all these studies have brought out interesting new features of the systems. BCS-BEC crossover, too, should not be any exception.

The simplest form of higher order many-body interactions would have been a three-fermion scattering. It has been argued by Holland et al. \cite{holland} that its effect will not be a prominent one, because in such a three-body interaction, the s-wave state is forbidden. The only three-body scattering there
could come from  p waves, which have very little contribution at sufficiently low temperatures. We, too, neglect such interactions for the time being. Instead, we shift our focus to a situation when, along the crossover path, some atom pairs have formed composite molecules, while some other pairs are yet to do so(they are still in the Cooper pair state). The newly formed bosons would scatter the pre-formed bosons (or, Cooper pairs), which is basically a three-body scattering. This interaction is sure to be important near the resonance point.  We take a variational mean field approach, and discuss the effect of this additional term  for all the four cases when either of the two-body and three-body interaction is attractive or repulsive.

\section*{II. EFFECTIVE COUPLING }
Here we start with  a two-species fermionic system. In addition to the fermion-fermion interaction ( denoted by $g_1$), and an additional  interaction ($g_2$) of the Feshbach variety which couples  a fermion of type $a$ with a $b$ fermion to form a bosonic molecule $B$, we also take into account  the scattering of pre-formed bosons or Cooper pairs by freshly formed bosons, strength of the interaction being given by $g_3$. The system is described by the Hamiltonian:

\begin{equation}
\begin{split}
\label{hamil}
 H=\sum(2\nu- \mu_B)B_0^{\dagger}B_0  +   \sum{\Tilde\epsilon_p^a a_p^{\dagger}a_p}  +  \sum{\Tilde\epsilon_p^b b_p^{\dagger}b_p} - g_1 \sum a_{p'}^{\dagger}b_{-p'}^{\dagger}b_{-p}a_p\\
  +  g_2 \sum[ B_0^{\dagger}a_{p}b_{-p} + {a^{\dagger}_{p}}b^{\dagger}_{-p}B_0 ]
  +g_3 \sum B_{q'}^{\dagger} a_{p'}^{\dagger}b_{-p'}^{\dagger}b_{-p}a_pB_q
  \end{split}
 \end{equation}
 
The ground state of the system is consequently given by \cite{rann, raka} a product wavefunction of the BCS ground state, and the ground state for the condensate part of the boson subsystem.
\begin{equation}
\label{grnd}
|\Psi\rangle = \prod (U_p+V_p a^{\dagger}_pb^{\dagger}_{-p})|0\rangle\otimes \mbox{exp}(-\alpha^2/2+\alpha B^{\dagger})|0\rangle 
\end{equation} 
where $\alpha = \sqrt{N_B}$, $N_B$ being the expectation value of the total number of bosons in the condensed state.
 
\indent From the Hamiltonian (\ref{hamil}) and the ground state wave function (\ref{grnd}), the ground state energy of the system would be
\begin{equation}
\begin{split}
E= \sum( \Tilde \epsilon_p^a+ \Tilde \epsilon_p^b) V_k^2 + g_1\sum_{p,p'} U_pV_pU_{p'}V_{p'} + (2\nu-\mu_B)\alpha^2 \\
+ 2g_2\alpha\sum_k U_pV_p+ g_3\alpha^2\sum_{p,p'} U_pV_pU_{p'}V_{p'} 
\end{split}
\end{equation} 

Differentiating $E$ with respect to $V_p$
\begin{equation}
\label{diff}
\begin{split}
4\epsilon^{+}_p V_p +2g_1\sum_{p,p'}U_{p'}V_{p'}(U_p-V_p^2/U_p) + 2g_2\alpha(U_p-V_p^2/U_p)\\
+2\alpha^2 g_3\sum_{p,p'}U_{p'}V_{p'}(U_p-V_p^2/U_p) =0
\end{split}
\end{equation} 
We assume that in the mean field framework, instead of couplings $g_1$, $g_2$ and $g_3$, this equation can be written in terms of an effective two-body coupling $g_{eff}$. This is in conformity with what Mora et al. \cite{mora} found out for a confined three-body problem : that it can be completely expressed in terms of two-body quantities. Equation (\ref{diff}) now takes the form -

\begin{equation}
4\epsilon^{+}_p V_p +2g_{eff}\sum_{p,p'}U_{p'}V_{p'}(U_p-V_p^2/U_p) =0
\end{equation}
The exact form of $g_{eff}$ is to be determined later. Drawing an analogy with the standard BCS treatment, we can write $-{g_{\mbox{eff}}\sum_{p}U_{p}V_{p} =\Delta}$, where $\Delta$ is the gap in the excitation spectrum.\cite{feyn}

Next, differentiating $E$ with respect to $\alpha$, we obtain
\begin{equation}
2\alpha(2\nu-\mu_B)+ 2 g_2\sum_k U_pV_p+ 2\alpha g_3\sum_{p,p'} U_pV_pU_{p'}V_{p'} =0
\end{equation}. Or,
\begin{equation}
\alpha = -\dfrac{g_2 \sum_{p} U_pV_p}{(2\nu -\mu_B + g_3\sum_{p,p'} U_pV_pU_{p'}V_{p'})}
\end{equation}
%\vspace{10 pt}
Putting $-{g_{\mbox{eff}}\sum_{p}U_{p}V_{p} =\Delta}$ as mentioned before,
\begin{equation}
\alpha = \dfrac{g_2\frac{\Delta}{g_eff}}{(2\nu -\mu_B + g_3\frac{\Delta^2}{{g_{eff}}^2})}
\end{equation}
Using this in equation (4), we get

\begin{equation}
\label{geff1}
\begin{split}
g_{eff}=g_1+g_3\alpha^2 - \dfrac{g_2^2}{(2\nu -\mu_B + g_3\frac{\Delta^2}{{g_{eff}}^2})}\\
=g_1 -\dfrac{g_2^2}{(2\nu -\mu_B + g_3\frac{\Delta^2}{{g_{eff}}^2})}+ \dfrac{g_3 g_2^2\frac{\Delta^2}{{g_{eff}}^2}}{(2\nu -\mu_B + g_3\frac{\Delta^2}{{g_{eff}}^2})^2}
\end{split}
\end{equation}

In the BCS limit, $\Delta$ goes as $e^{-\frac{1}{a_s}}$ \cite{leggett2}. Therefore, as $a_s\rightarrow 0$, $\Delta$ goes to zero much faster than $g_{eff}$ (which is proportional to $a_s$). Hence $\frac{\Delta^2}{g_{eff}^2}\rightarrow 0$ in this limit. 

We therefore have 
\begin{equation}
\label{geff0}
g_{eff}=g_1 -\frac{g_2^2}{(2\nu -\mu_B)}
\end{equation}

 If $g_1$, i.e, the two-fermion coupling is attractive, then this becomes $|g_{eff}|=|g_1| +\frac{g_2^2}{(2\nu -\mu_B)}$, a result analogous to the effective coupling in a standard two-species fermionic system, where only the fermion-fermion scattering is taken into account\cite{ohashi,levin1,raka}. This is consistent with the fact that in the extreme BCS limit, there are almost no composite bosons at all. Therefore, the Boson-Fermion scattering should not have any effect on the coupling parameter.

On the other hand, in the extreme BEC limit, $\Delta$ goes as $a_s^{-\frac{1}{2}}$\cite{leggett2}. Therefore, $\frac{\Delta^2}{g_{eff}^2} \approx \frac{1}{g_{eff}^3}$. Equation (\ref{geff1}) thus becomes:

\begin{equation}
\label{geff2}
\begin{split}
g_{eff}=g_1 - \dfrac{g_2^2}{(2\nu -\mu_B + \frac{g_3}{{g_{eff}}^3})}+ \dfrac{g_3 g_2^2}{{{g_{eff}}^3}(2\nu -\mu_B + \frac{g_3}{{g_{eff}}^3})^2}\\
=g_1 -\dfrac{g_2^2}{(2\nu -\mu_B + \frac{g_3}{{g_{eff}}^3})}\Big(1-\dfrac{\frac{g_3}{{g_{eff}}^3}}{2\nu -\mu_B + \frac{g_3}{{g_{eff}}^3}}\Big)\\
=g_1 - \dfrac{g_2^2(2\nu -\mu_B)}{(2\nu -\mu_B + \frac{g_3}{{g_{eff}}^3})^2}
\end{split}
\end{equation}
This is essentially a polynomial in degree 7, its real roots being the solutions for effective coupling.

We approximate the effective coupling in BCS side by equation (\ref{geff0}) and that of the BEC side by equation (\ref{geff2}) and study the crossover for various combinations of 2-body coupling $(g_1)$ and 3-body coupling $(g_3)$.
At resonance, it is known that $\Delta$ is proportinal to Fermi energy only \cite{chin}. So very near resonance, we treat $\Delta$ as a constant.\\\\

\section*{III. NEW BRANCHES IN THE CROSSOVER PICTURE : ALTERNATIVE PATHS }
\begin{figure}[h]
\includegraphics[scale=.8]{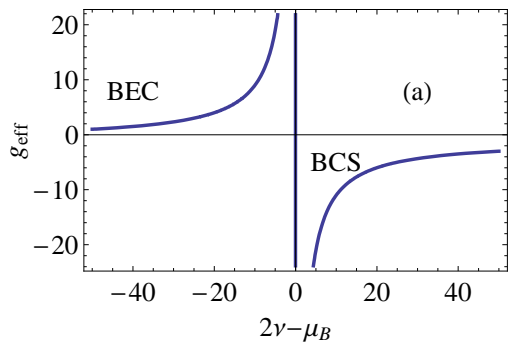}
\caption{(Color online) BCS-BEC Crossover}
\end{figure}

In the conventional BCS-BEC crossover picture, the two-body coupling ( and thus, the scattering length, too) is positive in the BEC side, and it goes to $\infty$ at resonance. It assumes a  negative value in the BCS side, and near resonance, goes to $-\infty$. 

In order to study the effects of three-body processes, here we assign some arbitrary values to $g_1$, $g_2$ and $g_3$ ( all the couplings are scaled by the Fermi energy $\epsilon_F$), and calculate the effective two-body couplings. $g_{eff}$ is plotted against the detuning $2\nu-\mu_B$ (which is also scaled by $\epsilon_F$)  to obtain the crossover picture. We note that even if the values of $g_1$, $g_2$ and $g_3$ are changed, the general qualitative trend of the coupling vs. detuning curve remains the same.\\\\
\textbf{Case 1 : Two-body Interaction Attractive, Three-body Interaction Repulsive:}
\begin{figure}[h]
\includegraphics[scale=.8]{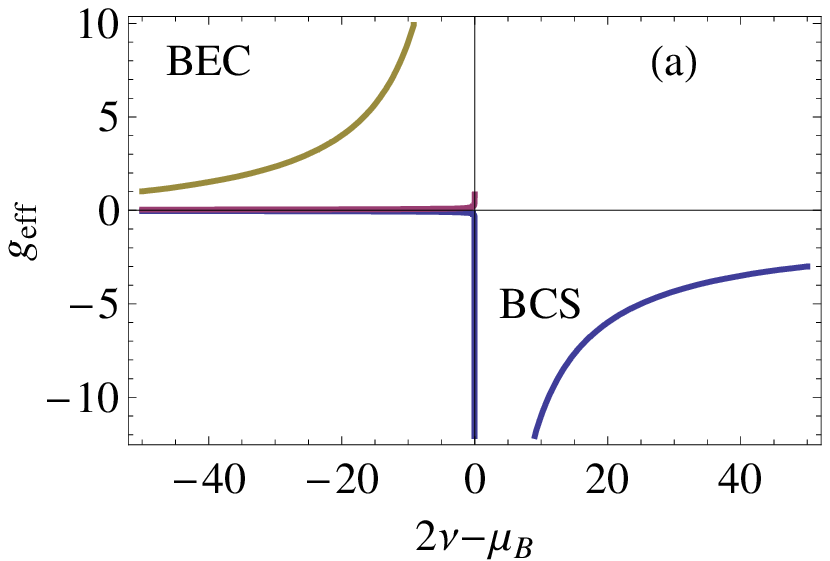}
\includegraphics[scale=.8]{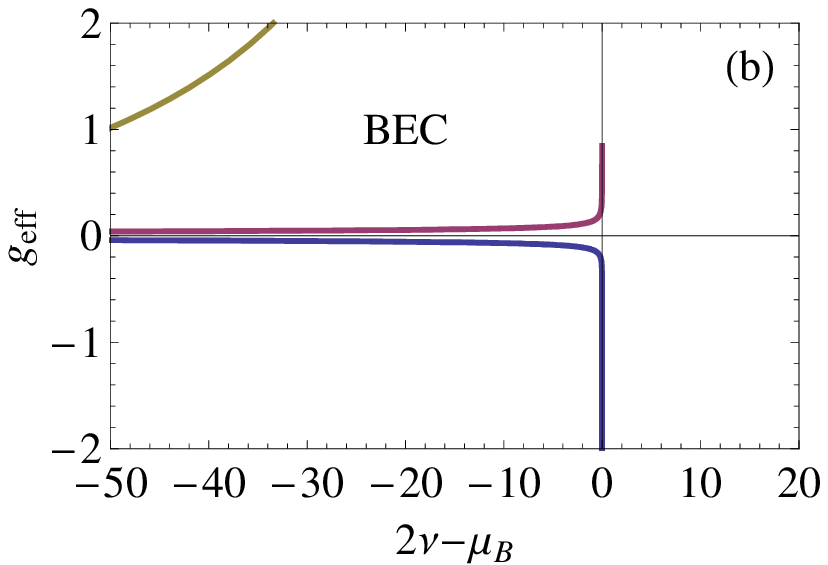}

\caption{(Color online) (a) Crossover paths near resonance when the two body interaction is attractive, and the three-body one  repulsive; (b) A closer view of the multiple roots at the resonance region. ( All parameters in Figs. 2-10 have been scaled by the Fermi energy.)}
\end{figure}

Here we choose $g_1=-1$, $g_2=10$ and $g_3= 0.1$ .The $g_{eff}$ vs. $2\nu-\mu_B$ curve is plotted in Figs 1(a), 1(b) and 2 for two different ranges. As shown in Figs 1(a) and 1(b),
when we approach $2\nu=\mu_B$, i.e, the resonance condition from the BEC side, we find there are two additional roots in addition to the $g_{eff}=\infty$ root. Therefore, when we move away slightly from the resonance towards the BEC region, there are three roots : one that corresponds to $g_{eff}$ in the absence of any $g_3$ ( and goes to $\infty$ at resonance), a negative root that goes almost linearly and reaches $g_1$ at resonance as evident from equation(\ref{geff2}), and a positive root that also varies almost in a linear fashion.
All the three roots survive even when the system is at deeper BEC domain.\\

\begin{figure}[h]
\includegraphics[scale=.8]{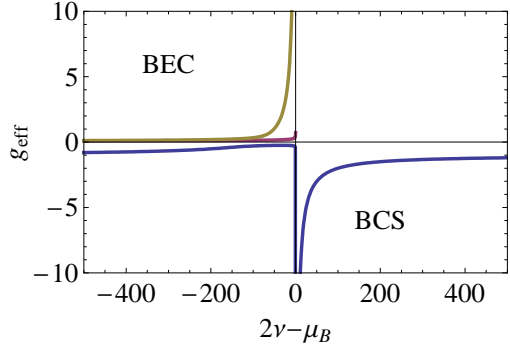}
\caption{(Color online) Crossover paths for a longer detuning range  when the two body interaction is attractive, and the three-body one is repulsive}
\end{figure}
In the BCS side, there is a single root which is identical to the effective coupling of the system in that region when there was no $g_3$. It goes to $-\infty$ at resonance.

Therefore, if starting from the BCS side, one tries to achieve the crossover, right after crossing the $2\nu=\mu_B$ point, one encounters three possible paths (Fig.2) -- one coming from $+\infty$, and the other two from finite values of $g_{eff}$. If the system takes either of the two lower  paths, the system actually bypasses the unitarity region ( where the scattering length diverges)in the BEC side. In such cases, although a region of unitarity will be present in the BCS side, its BEC side counterpart would be absent.\\\\

\textbf{Case 2 :  Two-body and Three-body Interactions Both Attractive:}\\

\begin{figure}[h]
\includegraphics[scale=.8]{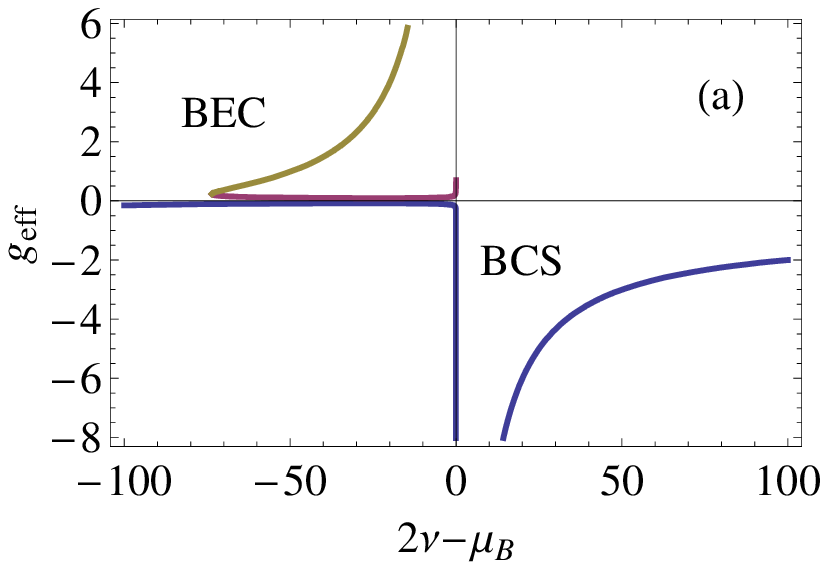}
\includegraphics[scale=.8]{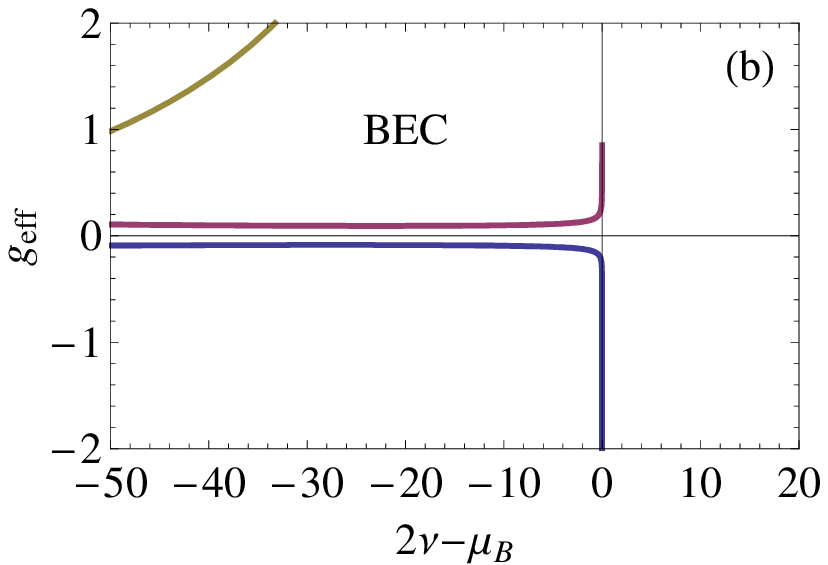}
\caption{(Color online) (a) Crossover paths near resonance when both the two body and three-body interactions are attractive ; (b) A closer view of the multiple roots at the resonance region.}
\end{figure}
Here we choose $g_1=-1$, $g_2=10$ and $-g_3=0.1$.
Here also, as shown in Figs. 3(a), 3(b),4, we have three real roots in the BEC side and a single real root on the BCS side.

In the BEC side, therefore, the system has the option to take any of the routes : one of them resembles the case of $g_3 =0$, and unitarity regions are there at both sides of the resonance. However, the other two branches can lead to a crossover scenario where again the unitarity is avoided in the BEC region.\\\\
\begin{figure}[h]
\includegraphics[scale=.8]{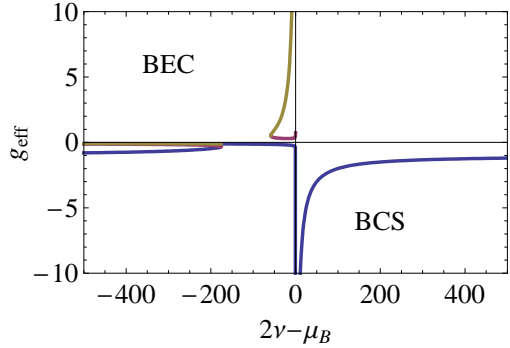}
\caption{(Color online) Crossover paths for a longer detuning range when both the two body and three-body interactions are attractive}
\end{figure}
We observe from Figs. 3(a) and 4 that that the upper two branches (including the traditional branch that we have for $g_3=0$) have a region of discontinuity, while the lowermost route is a continuous one.\\

\textbf{Case 3 :  Two-body and Three-body Interactions Both Repulsive:}\\
\begin{figure}[h]
\includegraphics[scale=.8]{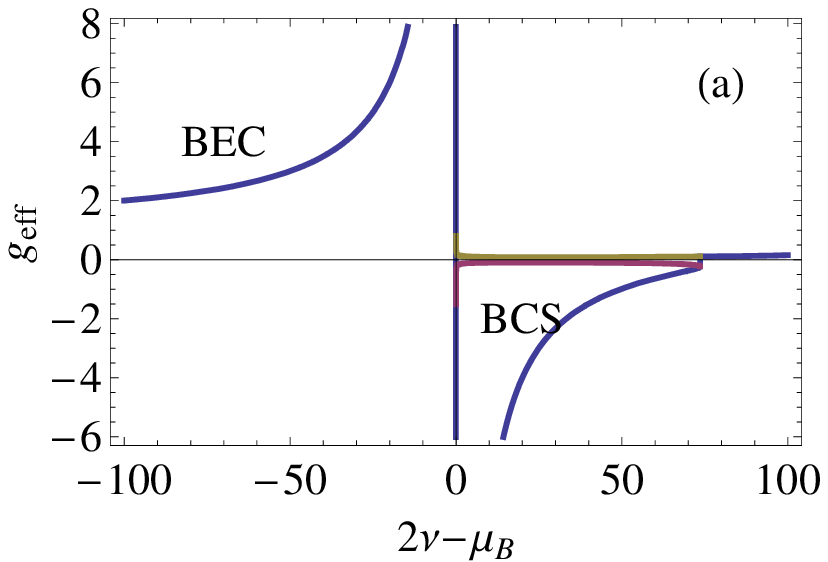}
\includegraphics[scale=.8]{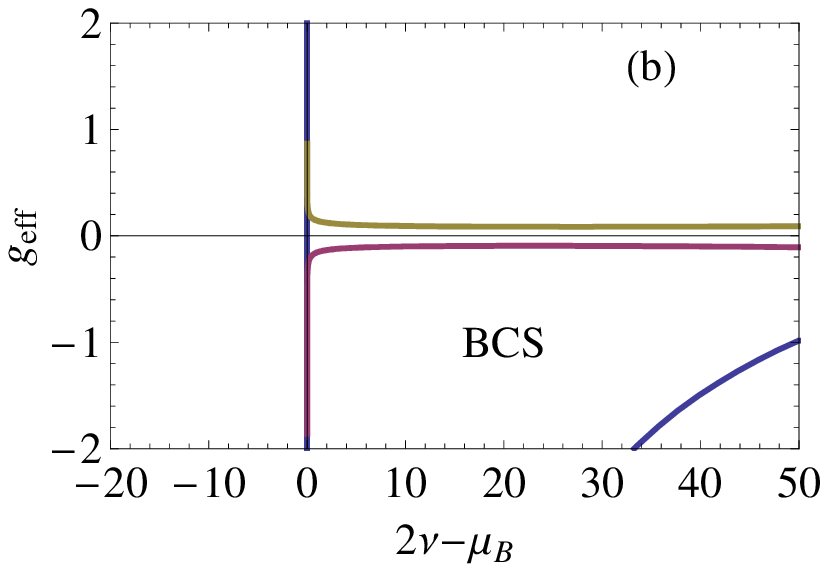}
\caption{(Color online)(a) Crossover paths near resonance when both the two body and three-body interactions are repulsive ; (b) A closer view of the multiple roots at the resonance region.}
\end{figure}

We choose $g_1=1$, $g_2=10$ and $g_3=0.1$. 
Here, as shown in Figs 5(a) and 5(b), there are three real roots in the BCS side, and one single root in the BEC region. \\

\textbf{ Case 4 : Two-body Interaction Repulsive, Three-body Interaction Attractive:}
Here we choose $g_1=1$, $g_2=10$ and $g_3=-0.1$.

\begin{figure}[h]
\includegraphics[scale=.8]{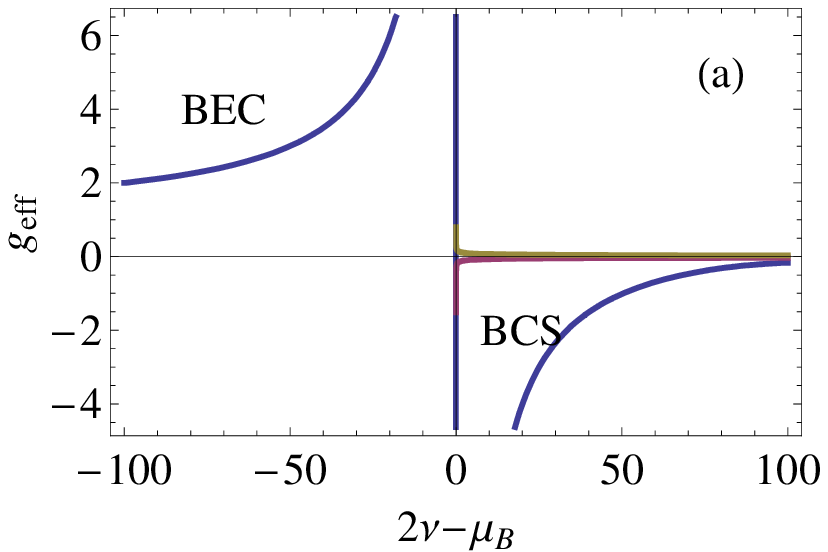}
\includegraphics[scale=.8]{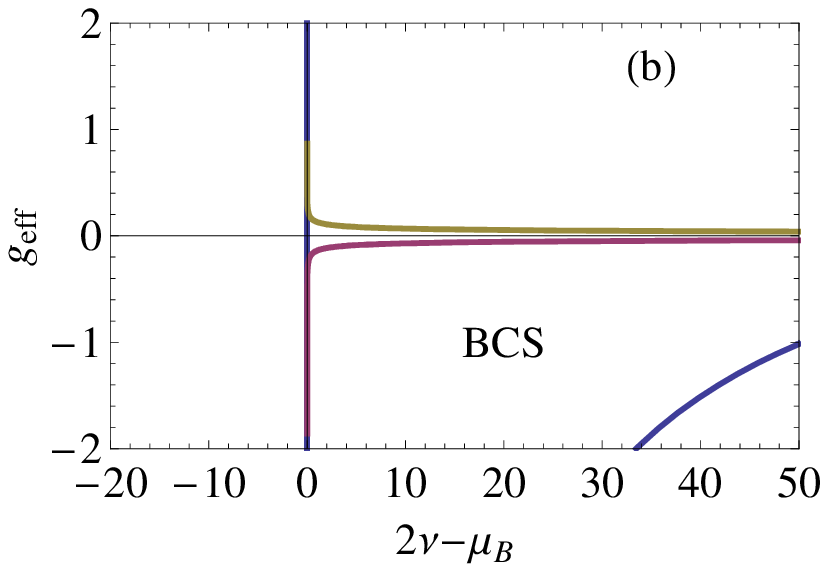}
\caption{(Color online)(a)Crossover paths near resonance when the two body interaction is repulsive, and the three-body one is attractive; (b) A closer view of the multiple roots at the resonance region.}
\end{figure}
In this case, as seen from Fig.6 the crossover picture closely resembles the earlier case, i.e, there are three possible routes in the BCS side, while a single route is available in the BEC side.\\\\

\section*{IV. WHICH PATH IS MORE FAVORABLE? ENERGY CONSIDERATIONS}
Let $E$ be the total energy of the system minus kinetic energy of the fermions. Now, if we compute and compare $E$ for all three branches, the branch corresponding to the minimum energy should be the one that the system favours.
At resonance we can treat $\Delta$ as constant. Using this, we plot $E$ with $2\nu-\mu_B$ and $g_{eff}$ (Fig. 7). In the BEC side, the lowermost branch corresponds to the minimum energy. Now, since this branch is associated with an attractive effective interaction, the BEC state should either collapse, or be a metastable one. Here we would get the latter, as the attraction is very weak.

\begin{figure}[h]
\includegraphics[scale=.7]{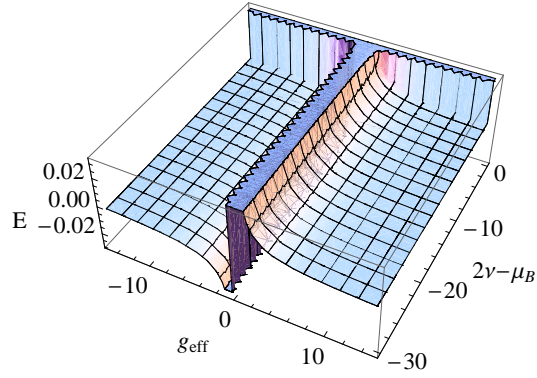}
\caption{(Color online) Variation in energy in the BEC side with effective coupling and detuning}
\end{figure}

Thus, we can have a striking property of the BCS-BEC crossover : If we start from a stable BEC state, we can achieve the BEC domain via Feshbach resonance; but if we start from the BCS side instead, we reach at a metastable BEC state. Thus the process is not totally reversible.

\begin{figure}[h]
\includegraphics[scale=.7]{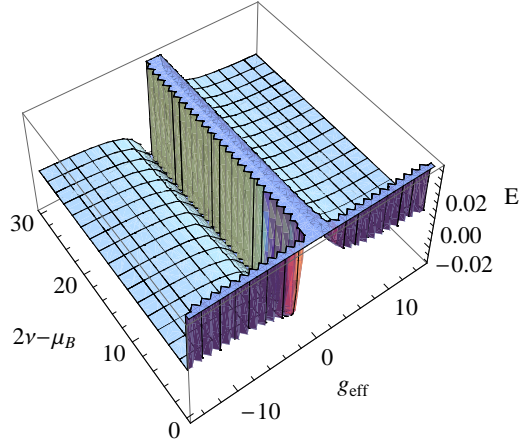}
\caption{(Color online) Variation in energy in the BCS side with effective coupling and detuning : gap is taken to be constant}
\end{figure}
In the BCS side, as apparent from the figure(9), the branch closer to $g_{eff} =0$ should be the favoured one. But in that case, the scattering length does not go to negative infinity, and the system cannot achieve the crossover. This is in contradiction with the well-established theoretical and experimental results \cite{leggett, rand1, zw1, zw2}. 

So we modify $\Delta$ using the form of the gap for weak coupling BCS \cite{pethick} :
\begin{equation}
\Delta_0 = 2\omega_c \mbox{exp}(-\frac{1}{\rho(0)g_{eff}})
\end{equation}
here $\rho(0)$ is the density of states in the Fermi level, and $\omega_c$ is the cutoff frequency for BCS model.
Thus we no longer treat the gap as a constant, and incorporate the coupling-dependence in it. This is justified, since the upper two branches are at the weak coupling domain, so the corresponding $\Delta$ should follow the expression for the BCS gap. Now the energy surface takes the form of Fig. 10.
\begin{figure}[h]
\includegraphics[scale=.7]{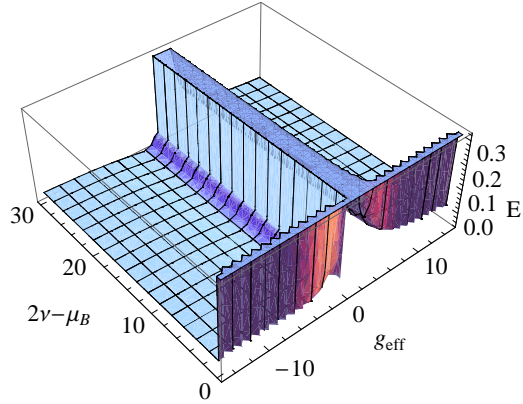}
\caption{(Color online) Variation in energy in the BCS side with effective coupling and detuning : gap is coupling dependent}
\end{figure}
It shows that in the negative $g_{eff}$ side, the energy is less as one goes away from $g_{eff}=0$. Therefore, the lowermost branch in the BCS side in figure(10) will be the favoured one.

\section*{V. CONCLUSION}
Here we have studied the BCS-BEC crossover in the presence of an additional three-body interaction term : the scattering of a Cooper pair by a newly formed boson near the resonance point. This has led to alternative crossover routes, and hence, brought out interesting  properties of the crossover phenomenon.  Most noteworthy of them is the non-reversibility of the process. If the two-body interaction is attractive (irrespective of whether the three-body interaction is attractive or repulive),then starting from a stable BEC system the BCS state can be reached via Feshbach resonance, but the path cannot be reversed : a start from the BCS side can only end up in a metastable BEC state (and not the stable one). This, we believe is an important finding and it reconfirms the need of more theoretical and experimental investigations along this line.

BEC-BCS crossover has been  experimentally investigated so far in $^6$Li and $^{40}$K systems. In most cases \cite{Bart,bourd, zw1, zw2} the system is first prepared in the BEC state, and the magnetic field is varied to obtain the BCS. This is completely  in agreement with our results.  In contrast, in  the experiment by C.A. Regal et al., the ultracold K-40 system went through the crossover in the opposite direction, i.e, it was a BCS-BEC crossover, and not a BEC-BCS one. Whether the final state here is a stable BEC or a metastable one with a weak attractive interaction (as predicted by our calculations) can be ascertained only after a study of the long-time response of the system. Moreover,  in this  experiment they first lowered the magnetic field slowly $(10 ms/G)^{-1}$ to bring it near resonance, and  then rapidly changed it $(50 \mu s/G)^{-1}$ to lower it further and obtain the BEC. Thus, although the crossover from the BCS to the resonance is an adiabatic one, it is a fast quench which takes the system from resonance to the BEC domain. So the mechanism may not have  followed the dynamics of a smooth crossover, and that could have resulted in a different final state ( a stable BEC state, for example). An experiment which probes the crossover starting from the BCS side, and changes the magnetic field adiabatically all through, might clarify this point. The final BEC state should also be studied carefully for a longer time so that a distinction between a metastable and a stable state can be easily made.

%%%%%%%%%%%%%%%%%%%%%%%%%%%%%%%%%%%%%%%%%%%%%%%%%%%%%%%%%%%%%%%%%%%%%%%%%%%%%%%%%%%%%%%%%%%%%%%%%%%%%%%%%%%%%%%%%%%%%%%%
\section*{ACKNOWLEDGEMENT}
The author would like to thank her supervisor Prof.J.K. Bhattacharjee for his guidance, support and constructive suggestions at various stages of this work.
%%%%%%%%%%%%%%%%%%%%%%%%%%%%%%%%%%%%%%%%%%%%%%%%%%%%%%%%%%%%%%%%%%%%%%%%%%%%%%%%%%%%%%%%%%%%%%%%%%%%%%%%%%%%%%%%%%%%%%%%

\end{document}